# Interpretation in Quantum Physics as Hidden Curriculum

Charles Baily and Noah D. Finkelstein

*Department of Physics, University of Colorado, Boulder, CO 80309, USA*

**Abstract.** Prior research has demonstrated how the *realist perspectives* of classical physics students can translate into specific beliefs about quantum phenomena when taking an introductory modern physics course. Student beliefs regarding the interpretation of quantum mechanics often vary by context, and are most often in alignment with instructional goals in topic areas where instructors are explicit in promoting a particular perspective. Moreover, students are more likely to maintain realist perspectives in topic areas where instructors are less explicit in addressing interpretive themes, thereby making such issues part of a *hidden curriculum*. We discuss various approaches to addressing student perspectives and interpretive themes in a modern physics course, and explore the associated impacts on student thinking.

## INTRODUCTION

In physics education research, the term *hidden curriculum* generally refers to those aspects of science and learning about which students maintain or develop attitudes and opinions, but which are primarily only implicitly addressed by instructors. [1] Students may hold varying beliefs regarding the relevance of course content to real-world problems, the coherence of scientific knowledge, or even the purpose of science itself, depending (in part) on the choices and actions of their instructors. Prior research has demonstrated that student attitudes regarding such matters tend to remain or become less expert-like when instructors are not explicit in addressing them. [1]

The role of interpretation in science can become particularly significant when more than one physical model is successful in accounting for a set of experimental results – practicing physicists may favor one type of model over another based on many factors, such as physical intuition or simplicity. Quantum mechanics has been plagued by questions of interpretation from the beginning, and the physical interpretation of quantum theory has been historically considered a matter of philosophical taste, and not subject to experimental verification. However, the theoretical work of Bell [2] and the more recent onset of "single-quanta" experiments [3, 4] have shown that some interpretive themes from quantum mechanics (e.g., determinacy vs. indeterminacy, locality vs. non-locality) can be put to experimental test. The growth in quantum information theory and experiment has made the physical interpretation of quantum mechanics more relevant than ever to practicing physicists. [5]

When considering student perspectives on quantum phenomena, it is important to understand that through instruction in classical physics, or even from everyday experience, many introductory students develop *realist perspectives*, based partly on intuitive conceptions of particle and wave phenomena. A realist perspective would be deterministic, in the sense that physical quantities (such as the position or momentum of a particle) are assumed to be objectively real (i.e. observation independent), and when specified can be accurately predicted for all future times. For introductory quantum physics students, realist perspectives may translate into specific beliefs about quantum phenomena: e.g. quanta are always localized in space; or, that the probabilistic nature of quantum mechanics is a consequence of classical ignorance, as opposed to a more fundamental indeterminacy.

From a number of post-instruction interviews with students from several introductory modern physics courses recently taught at the University of Colorado, we find that students develop attitudes and opinions regarding specific interpretive themes in quantum physics, regardless of whether and how those themes had been addressed by their instructors. Our research has also shown that, the less explicit an instructor is in addressing student perspectives within a given topic area, the greater the likelihood for students to favor realist perspectives within that specific context. [6] In other words, the less student perspectives on quantum mechanics are explicitly addressed by instructors, the more they become part of a hidden curriculum. In this paper we explore how this hidden curriculum may (or may not) be addressed by modern physics instructors, by first examining the impact of specific instructional approaches on student thinking; we then summarize results from a more refined characterization of student perspectives [7] in order to better understand the complex relationship between instructors, students and educational practices.

## INSTRUCTIONAL APPROACHES IN MODERN PHYSICS COURSES

In this section, we discuss not only the types of quantum interpretations favored by instructors, but also their approaches to addressing student perspectives in an introductory modern physics course. Our characterizations of instructor stances on interpretive themes in quantum mechanics are based on classroom observations, an analysis of course materials, and interviews with instructors, and have been described in prior work; [6] these approaches can be best illustrated by how each instructor addressed the double-slit experiment. *Realist/Statistical* instructors taught that each particle passes through one slit or the other, but that it is impossible to determine which one without destroying the interference pattern. *Matter-Wave* instructors promoted a wave-packet description of individual quanta, where each electron propagates as a delocalized wave through both slits, interferes with itself, and then becomes localized upon detection. *Copenhagen* instructors said that a *quantum mechanical wave of probability* passes through both slits, but that posing which-path questions will disrupt the interference pattern; *Agnostic* instructors were similar, but emphasized predicting features of the interference pattern (mathematical calculation) over questions of interpretation.

We describe here four specific approaches to addressing interpretation in four different modern physics courses recently taught at the University of Colorado, and demonstrate significant differences in student thinking associated with these approaches. **Figs. 1** & **2**, where letters refer to specific instructors, show aggregate student responses to two items from an end-of-term online survey, illustrating both the differential impacts of these instructional approaches, as well as the mixed nature of student responses across contexts. **Fig. 1** contains student responses to an essay question on interpretations of the double-slit experiment with single quanta. In this topic area, instructors had been explicit in teaching one particular interpretation, *though not explicitly as an interpretation*; student responses in this context were generally reflective of the teaching goals for each course. **Fig. 2** shows how these same students responded to the statement: *An electron in an atom has a definite (but unknown) position at each moment in time.* Disagreement could be consistent with either a Quantum (wave-packet) or a Copenhagen/Agnostic perspective, whereas agreement would be more consistent with a Realist perspective. Instructors from three of the four courses paid considerably less attention to interpretive themes at later stages of the course, as when students learned about the Schrödinger model of hydrogen. As seen in **Fig. 2**, students from every course were more likely to agree with this statement than disagree, including students from the Matter-Wave courses.

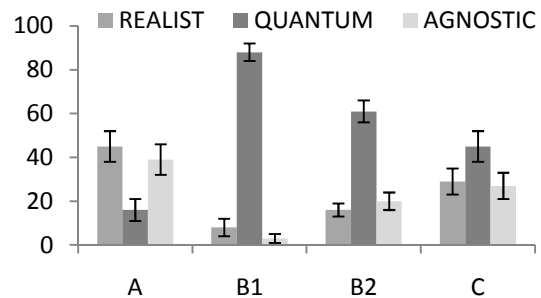

**FIGURE 1.** Post-instruction student responses (in percent) to an essay question on interpretations of the double-slit experiment, from four modern physics courses using different instructional approaches. [A = Realist/Statistical; B1 & B2 = Matter-Wave; C=Copenhagen/Agnostic, as described in the text]. Student responses (Realist, Quantum, Agnostic) are as described in the text. Error bars represent the standard error on the proportion; N ~ 100 for each of the four courses.

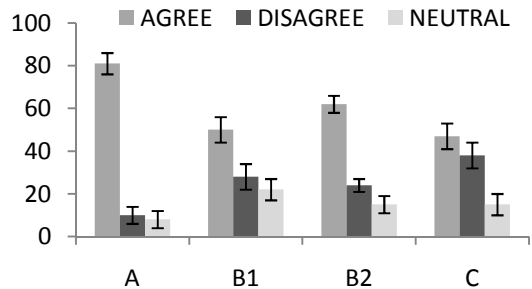

**FIGURE 2.** Post-instruction student responses (in percent) from four modern physics courses of different instructional approaches [A = Realist/Statistical; B1 & B2 = Matter-Wave; C = Copenhagen/Agnostic, as described in the text] to the statement: *An electron in an atom exists at a definite (but unknown) position at each moment in time.* Error bars represent the standard error on the proportion; N ~ 100 for each of the four courses.

# Specific Instructional Practices

We provide here a more detailed discussion of the specific instructional approaches employed in the courses described above. Letters refer to specific instructors, as given in the figure captions.

**A. Teach an interpretation that aligns with student intuition, but is less favored by practicing physicists, without discussing alternatives:** Instructor A taught from a Realist/Statistical perspective (though he did not call it such), and explicitly referred to this as his own interpretation of quantum phenomena, one that other physicists would not necessarily agree with. Beyond the double-slit experiment, students were also explicitly taught to think of atomic electrons as localized particles, and that energy quantization is the result of the average behavior of atomic particles. There was no discussion of alternatives to the perspective being promoted in class. Student responses from this course in both contexts could be considered in alignment with Instructor A's explicit learning goals: they were the most likely to prefer a Realist interpretation of the double-slit experiment [each electron goes through either one slit or the other, but not both], as well as the most likely to agree that atomic electrons exist as localized particles. We believe student responses from this course are reflective not only of explicit instruction, but also that this particular kind of interpretation of quantum mechanics is in agreement with realist expectations.

**B1. Teach one interpretation (though not explicitly as an interpretation) in some topic areas (particularly at the beginning of the course) and expect students to generalize to other contexts on their own:** When first teaching this modern physics course, Instructor B was explicit in modeling single quanta in the double-slit experiment as delocalized waves that pass through both slits simultaneously, though he did not frame this discussion in terms of modeling or interpretation, but rather as a fact that students needed to incorporate into their understanding. Students from this Matter-Wave course overwhelmingly preferred a wave-packet (*Quantum*) description of individual electrons [each electron passes through both slits simultaneously and interferes with itself]. However, these students did not seem to generalize this notion of particles as delocalized waves in the double-slit experiment to the context of atoms, with a majority still agreeing that atomic electrons exist as localized particles. Students were more likely to retain realist notions in a topic area where Instructor B was not explicit regarding interpretation.

**B2. Teach one interpretation (though not explicitly as an interpretation) in some topic areas, combined with a more general discussion of interpretative themes towards the end of the course:** Instructor B later taught a second modern physics course in a similar manner, but this time devoted two days of lecture time near the end of the course to interpretive themes in quantum mechanics, including a discussion of the interpretive aspects of the double-slit experiment, but without reference to atomic systems. Student responses were similar to the previous Matter-Wave course (B1) on interpretations of the double-slit experiment, but a majority of students still preferred a Realist stance on atomic electrons.

**C. Teach a Copenhagen/Agnostic perspective, or de-emphasize questions of interpretation:** Instructor C felt that introductory students do not have the requisite sophistication to appreciate the nuances of interpretive issues in quantum mechanics. And though he did touch on such themes during the course, he ultimately emphasized a perspective that is more pragmatic than philosophical, as when faced with the student question of whether particles have a definite but unknown position, or have no definite position until measured:

"Newton's Laws presume that particles have a well-defined position and momentum at all times. Einstein said that we can't know the position. Bohr said, philosophically, it has no position. *Most physicists today say: We don't go there. I don't care as long as I can calculate what I need.*"

Student responses from this course regarding the double-slit experiment were more varied than with others – students were not only likely to prefer an *Agnostic* stance [quantum mechanics is about predicting the interference pattern, not discussing what happens between], a significant number of students (30%) preferred a Realist interpretation – more than with the Matter-Wave courses, but less so than with the Realist/Statistical course. Nearly half of all students from this course also preferred a Realist stance on atomic electrons.

# REFINING CHARACTERIZATIONS OF STUDENT PERSPECTIVES

Results from a more detailed exploration of student perspectives [7] provide insight into the nuanced and contextual nature of student responses, as well as the tendency among students to prefer realist stances on quantum phenomena.

Slightly more than half of the students from our interviews (ten of nineteen) demonstrated a preference for realist interpretations of quantum mechanics. However, the nature of these students' realist perspectives were not necessarily of the character we had anticipated from earlier studies. Only three of the seven students who preferred local, realist interpretations of quantum physics expressed confidence in the correctness of their perspectives, whereas four others differentiated between what made intuitive sense to them (Realist) and what they perceived as a correct response (Quantum). In addition to *splits* between intuition and authority, some of the seemingly contradictory responses from students may also be explained by a preference for a *mixed ontology* (a pilot-wave interpretation, wherein quanta are simultaneously *both* particle *and* wave). The realist beliefs of these three students were of a decidedly non-local character: localized quantum entities follow trajectories determined by the interaction of non-local quantum waves with the environment. The perspectives of students expressing these types of beliefs (quanta as simultaneously wave and particle) were at odds with how wave-particle duality was addressed in class by their instructors (i.e., quanta are sometimes described by waves, and sometimes as particles, but never both simultaneously).

We also find it significant that almost every interviewed student expressed distaste for deterministic ideas in the context of quantum physics, although it had been anticipated that Realist/Statistical students might favor such notions. Not only did most students say they were unfamiliar with the word *determinism* within the context of physics, practically every student believed either that the behavior of quantum particles is inherently probabilistic, or that the Uncertainty Principle places a fundamental limit on human knowledge of quantum systems, or a combination of both stances. Students who preferred realist perspectives in the interviews were most likely to favor the latter stance. A realist and a probabilistic perspective are not necessarily in conflict – favoring both can be indicative of how students do not distinguish between classical ignorance and the more fundamental uncertainty associated with quantum measurements. Probabilistic descriptions of the outcomes of measurements are necessary when knowledge of the initial conditions is incomplete.

Due to the limited number of participants, no connection could be discerned in these interviews between each student's preferred perspective and the specific approach taken by their instructor.

## CONCLUSIONS

In exploring student perspectives on quantum physics, we find it natural that students would have attitudes regarding some interpretive themes, in that we were ultimately probing each student's ideas about the very nature of reality, and the role of science in describing it: Is the universe deterministic or inherently probabilistic? When is a particle a particle, and when is it a wave, and what is the nature of this wave? Is it unscientific to discuss the unobservable? We find that students, as a form of sense-making, develop ideas and opinions regarding the interpretation of quantum mechanics, regardless of how their instructors addressed matters of interpretation in class.

Questions of interpretation in quantum mechanics are of both personal and academic interest to students, and modern physics instructors should recognize the potential impact on student thinking when choosing to de-emphasize interpretation in an introductory course. Moreover, interpretation is a significant aspect of scientific thinking, and students should benefit from not only understanding how to make use of equations, but also to interpret physical meaning from those equations (as well as the individual terms that make up those equations). Although many instructors may argue that introductory students do not have the requisite sophistication to appreciate matters of interpretation in quantum mechanics, we believe interpretive discussions may be incorporated into most any topic area in physics (for example, which is more fundamental [real]: the electric field or the electric potential?). Questions of interpretation may also be addressed in terms of scientific modeling, or Nature of Science issues, aspects of epistemological sophistication that are often emphasized in physics education research as an explicit goal of instruction.

## ACKNOWLEDGMENTS

This work was supported in part by NSF CAREER Grant No. 0448176 and the University of Colorado.